\documentclass[aps,prd,reprint,superscriptaddress,nofootinbib]{revtex4-1}

\usepackage{amssymb,amsmath,bm,natbib}
\usepackage[dvipsnames]{xcolor}
\usepackage[colorlinks = true,
            linkcolor = Maroon,
            urlcolor  = Maroon,
            citecolor = Maroon]{hyperref}
\usepackage{microtype}
\usepackage{graphicx}
\usepackage{xspace}
\usepackage{bm}

\begin{document}

\title{The Migdal effect in solid crystals and the role of non-adiabaticity}

\author{Angelo~Esposito}
\email{angelo.esposito@uniroma1.it}
\affiliation{Dipartimento di Fisica, Sapienza Universit\`a di Roma, Piazzale Aldo Moro 2, I-00185 Rome, Italy}
\affiliation{INFN Sezione di Roma, Piazzale Aldo Moro 2, I-00185 Rome, Italy}

\author{Andrea~Rocchi}
\affiliation{Dipartimento di Fisica, Sapienza Universit\`a di Roma, Piazzale Aldo Moro 2, I-00185 Rome, Italy}

\date{\today}

\begin{abstract}
\noindent We systematically apply the Born--Oppenheimer approximation to show that the Migdal effect in a solid crystal is entirely due to non-adiabatic effects, namely the deviation of the wave function from exact factorization of the electronic and nuclear contributions. The matrix element obtained this way matches exactly the result found by means of a previously derived low energy effective theory.
\end{abstract}

\maketitle


\section{Introduction}

\noindent In a direct detection experiment, one attempts to observe dark matter in the lab by leveraging its possible non-gravitational interactions with our detectors. In particular, several searches aim at probing the coupling between dark matter and nuclei, by looking for elastic scattering events of the former off the latter~\cite[e.g.,][]{PandaX-4T:2021bab,DarkSide-50:2022qzh,LZ:2024zvo,XENON:2025vwd}. Nonetheless, when there is a large mismatch between the mass of the target nucleus and the mass of the dark matter one is looking for, the elastic scattering process becomes inefficient at converting the dark matter energy into nuclear recoil energy, thus falling below the experimental energy threshold. Consequently, the constraints that one obtains in the sub-GeV mass range are considerably weaker than those obtained for higher masses~\cite[e.g.,][]{SuperCDMS:2020aus,CRESST:2024cpr}.

One possibility to overcome this limitation is to look for {\it inelastic} processes~\cite[e.g.,][]{Essig:2011nj,Kouvaris:2016afs}, such as the so-called Migdal effect~\cite[e.g.,][]{Migdal1939,Migdal:1941,Ibe:2017yqa}. This is a process where the release of energy and momentum to a nucleus induces the excitation of some electronic degrees of freedom. A textbook example is the excitation of the electron of a Hydrogen-like atom to some higher energy level, following a kick received by its nucleus (cf. \S41 of~\cite{Landau:1991wop}). In recent years, the Migdal effect has received a great deal of attention as a promising avenue to look for dark matter--nucleus interactions even in the sub-GeV region. Intensive study has been undertaken both from the theory side~\cite{Ibe:2017yqa,Vergados:2005dpd,Moustakidis:2005gx,Bernabei:2007jz,Essig:2019xkx,Cox:2022ekg,Dolan:2017xbu,Bell:2019egg,Baxter:2019pnz,Liang:2019nnx,Liu:2020pat,Kahn:2020fef,Flambaum:2020xxo,Bell:2021zkr,Acevedo:2021kly,Wang:2021oha,Blanco:2022pkt,Liang:2020ryg,Knapen:2020aky,Liang:2022xbu,Berghaus:2022pbu,Li:2022acp,Bell:2023uvf,Yun:2023huf,Gu:2023pfg,Herrera:2023xun,Nakano:2024qkg,Maity:2024vkj} and from the experimental one~\cite{XENON:2019zpr,DarkSide:2022dhx,Adams:2022zvg,Xu:2023wev,SENSEI:2023zdf,Kahn:2024nyv}.

In this context, semiconductors are excellent targets, as their electronic bandgaps are $\mathcal{O}({\rm eV})$, thus allowing to probe dark matter masses as low as $m_\chi \sim \mathcal{O}({\rm MeV})$. The Migdal effect, in this case, would
consist in the promotion of an electron from the valence to the conduction band, following an interaction between dark matter and nuclei.

The theoretical description of this phenomenon for a generic dark matter mass is rather challenging, and was originally approached only within certain regimes of exchanged momenta~\cite{Liang:2020ryg,Knapen:2020aky,Liang:2022xbu}. By building upon the results of~\cite{Knapen:2020aky}, a general theory of the Migdal effect in semiconductors was presented in~\cite{Berghaus:2022pbu}, by developing an effective theory which takes advantage of the large separation between the typical energies of the electronic excitations (few eVs) and those of the vibrational modes of the crystal (up to hundreds of meVs). This theory allows to bypass the intricacies of the microscopic nuclear interactions, extending the calculation of the Migdal rate to all dark matter masses. These results have indeed been used by the SENSEI collaboration to put the currently most stringent bounds on dark matter--nucleus interactions at masses down to the MeV~\cite{SENSEI:2023zdf}.

In spite of this, the theoretical understanding of the Migdal effect in semiconductors might still be incomplete. In fact, it was highlighted in~\cite{Blanco:2022pkt} that, at least in the context of a molecular target, the total Migdal event rate gets contributions from two different types of effects: so-called adiabatic and non-adiabatic. The adiabatic contribution is due to the mismatch between the nuclear coordinates, to which the dark matter couples, and the coordinates of the actual degrees of freedom in the molecular center-of-mass frame, very much analogous what causes the Migdal effect for a single atomic target (see again \S41 of~\cite{Landau:1991wop}). The non-adiabatic contribution, instead, is due to the coupling between nuclei and electrons, ultimately due to the response of the electrons to a displacement of the nuclei. In~\cite{Blanco:2022pkt}, it was also noticed that this non-adiabatic contribution might dominate over the adiabatic one, for sufficiently large molecules.

In the effective theory presented in~\cite{Berghaus:2022pbu} there is no clear separation between adiabatic and non-adiabatic contributions, which thus leaves one wondering: does the effective theory capture all possible effects? And, if not, should we revisit the current experimental bounds?

In this work, we implement a systematic version of the Born--Oppenheimer approximation to compute the amplitude for the Migdal event in a formally infinite crystal. In analogy with~\cite{Blanco:2022pkt}, we are able to clearly disentangle the adiabatic and non-adiabatic contributions. We show that the first one vanishes, while the second one exactly reproduces what found in~\cite{Berghaus:2022pbu}, thus confirming the validity of the results of the effective theory.

\vspace{0.5em}

\noindent {\it Conventions:} We employ natural units, $\hbar = c = 1$.


\section{The Born--Oppenheimer approximation, done systematically}

\noindent We start by reviewing the systematic implementation of the Born--Oppenheimer approximation, following closely the approach taken in~\cite{born1996dynamical}. The general Hamiltonian of the nuclei and electrons in a solid is given by,
\begin{align} \label{eq:Hsolid}
    \begin{split}
        H(\bm X,\bm x) ={}& - \sum_I \frac{\nabla^2_{\bm X_I}}{2M_I}  - \sum_i \frac{\nabla^2_{\bm x_i}}{2m} + V\big(\bm X, \bm x\big)  \\
        \equiv{}& - \frac{m}{M} \sum_I \frac{M}{M_I} \frac{\nabla^2_{\bm X_I}}{2m} +  H_e(\bm X,\bm x) \,, 
    \end{split}
\end{align}
where $M_I$ and $m$ are the masses of nuclei and electrons, and $\bm X$ and $\bm x$ their coordinates. For reasons that will soon be clear, we also introduced a mass $M$, which we take to be comparable to all the nuclear masses, $M \sim M_I$. 

Let us stress an important aspect right away. In a formally infinite system, such as an idealized solid, the $\bm X$ coordinates appearing in the equation above coincide with the actual positions of the nuclei, to which the dark matter couples to. This is because, contrary to a finite system, the degree of freedom corresponding to the center-of-mass is infinitely heavy, and thus plays no role. This will be crucial for the argument that we present in the next section.

The Hamiltonian in Eq.~\eqref{eq:Hsolid} defines the Schr\"odinger equation for the full system, which is described by a wave function, $\Psi_\alpha( \bm X,\bm x)$, satisfying,
\begin{align}
    H(\bm X,\bm x) \, \Psi_\alpha(\bm X,\bm x) = E_\alpha \, \Psi_\alpha(\bm X,\bm x) \,,
\end{align}
where $\alpha$ is a set of quantum numbers labeling the state of the full system.
Together with this, one also considers an electronic Schr\"odinger equation, which is defined to be,
\begin{align} \label{eq:electronic}
    H_e(\bm X,\bm x) \, \phi_n(\bm x|\bm X) = \mathcal{E}_n(\bm X) \, \phi_n(\bm x|\bm X) \,,
\end{align}
where $n$ represents a collection of quantum numbers labeling the electronic state.
The equation above is a differential equation for the dependence on the electronic coordinates, $\bm x$, and the nuclear coordinates, $\bm X$, only enter as parameters. Thus, the electronic wave functions and energies will inherit this parametric dependence.

Now, the Born--Oppenheimer approximation can be organized as a systematic expansion in the ratio between electronic and nuclear masses, encoded in the small parameter $\epsilon \equiv (m/M)^{1/4} \ll 1$. This appears both explicitly in the nuclear kinetic term and implicitly in an expansion of the nuclear coordinates in small displacements around their equilibrium values, $\bm X_I \equiv \bar{\bm X}_I + \epsilon \, \bm u_I$. 
Accordingly, one can expand all the relevant quantities in this parameter, 
\begin{align}
    \Psi_\alpha ={}& \Psi^{(0)}_\alpha(\bm u,\bm x) + \epsilon \, \Psi^{(1)}_\alpha(\bm u,\bm x) + \epsilon^2 \Psi^{(2)}_\alpha(\bm u,\bm x) + \dots \,, \notag \\
    \phi_n ={}& \phi^{(0)}_n(\bm x) + \epsilon \,\phi^{(1)}_n(\bm x|\bm u) + \epsilon^2 \phi^{(2)}_n(\bm x|\bm u) + \dots \,, \\
    \mathcal{E}_n ={}& \mathcal{E}_n^{(0)} + \epsilon^2 \mathcal{E}^{(2)}(\bm u) + \dots \,. \notag
\end{align}
Importantly for us, $\phi_n^{(0)}(\bm x)$ and $\mathcal{E}_n^{(0)}$ are independent on $\bm u$, $\phi_n^{(1)}(\bm x|\bm u)$ is linear in it, while $\phi_n^{(2)}$ and $\mathcal{E}_n^{(2)}$ are quadratic in it. The electronic eigenvalue has no linear term, since this is precisely what defined the equilibrium positions of the nuclei, i.e. $\mathcal{E}_n^{(1)}(\bm u) = 0$~\cite{born1996dynamical}.

Upon solving both the electronic and the full Schr\"odinger equation perturbatively in $\epsilon$~\cite{born1996dynamical}, one finds that, up to $\mathcal{O}(\epsilon^2)$, the full wave function can be written in a factorized way, $\Psi_{n\lambda}(\bm u,\bm x) \simeq \phi_n(\bm x|\bm u) \chi_{n\lambda}(\bm u)$. Here $\chi$ is a wave function describing the nuclear state, labeled by quantum numbers $\lambda$. It is also a solution to a Schr\"odinger equation, perturbatively solved in $\epsilon$. Its expression beyond leading order, however, is considerably more involved~\cite{born1996dynamical}, and we do not report it here. At lowest order, it reads,
\begin{align} \label{eq:harmonic}
    \left[- \sum_I \frac{M}{M_I} \frac{\nabla^2_{\bm u_I}}{2m} + \mathcal{E}_n^{(2)}(\bm u) \right] \chi_{n\lambda}^{(0)}(\bm u) = E_{n\lambda}^{(2)} \, \chi_{n\lambda}^{(0)}(\bm u) \,.
\end{align}
Since $\mathcal{E}_n^{(2)}(\bm u)$ is quadratic in the displacement, this is nothing but an harmonic oscillator. 

The sharp separation highlighted above between the electronic and nuclear problems breaks down at $\mathcal{O}(\epsilon^3)$, where the wave function cannot be written in a factorized fashion anymore. The first correction beyond factorization can be found explicitly, and the complete wave function is given by~\cite{born1996dynamical},
\begin{widetext}
\begin{align} \label{eq:Psi}
    \begin{split}
        \Psi_{n\lambda}(\bm u,\bm x) ={}& \left[ \phi_n^{(0)}(\bm x) + \epsilon \, \phi_n^{(1)}(\bm x|\bm u) + \epsilon^2 \, \phi_n^{(2)}(\bm x|\bm u) + \epsilon^3 \, \phi_n^{(3)}(\bm x|\bm u) \right] \left[ \chi_{n\lambda}^{(0)}(\bm u) + \epsilon \, \chi_{n\lambda}^{(1)}(\bm u) + \epsilon^2 \, \chi_{n\lambda}^{(2)}(\bm u) + \epsilon^3 \, \chi_{n\lambda}^{(3)}(\bm u) \right] \\
        & \; + \frac{\epsilon^3}{m} \sum_I \bm\nabla_{\bm u_I}\chi_{n\lambda}^{(0)}(\bm u) \cdot \sum_{m\neq n} \frac{\big\langle \phi_m^{(0)}\big| \bm \nabla_{\bm u_I}\phi_n^{(1)}(\bm u)\big\rangle}{\mathcal{E}_m^{(0)}- \mathcal{E}_n^{(0)}} \phi_m^{(0)}(\bm x) + \mathcal{O}(\epsilon^4) \\
        \equiv{}& \phi_n(\bm x|\bm u) \chi_{n\lambda}(\bm u) + \delta \Psi_{n\lambda}(\bm u,\bm x) + \mathcal{O}\big(\epsilon^4\big) \,,
    \end{split}
\end{align}
\end{widetext}
where the expansion of the first line must be truncated at third order. 
The first factorized term, $\phi\,\chi$, is the traditional {\it adiabatic} part of the wave function, while the second, $\delta \Psi$, is the first correction to this factorization, and it is typically dubbed as {\it non-adiabatic} correction. The matrix element in the numerator of $\delta \Psi$ is evaluated as an integral over just the electronic coordinates, $\bm x$.
As one can see, starting at this order, different electronic states contribute to the total wave function. 

\vspace{1em}

Finally, we stress that the perturbative approach just described breaks down for sufficiently large nuclear displacements. Indeed, as seen in Eq.~\eqref{eq:harmonic}, the lowest order nuclear wave function is Gaussian, with a spread of order of the Bohr radius, $a$, as shown, for example, in~\cite{weinberg2015lectures}. The other quantities appearing in $\delta \Psi$ scale as~\cite{weinberg2015lectures},
\begin{align}
    \mathcal{E}_n^{(0)} \sim \frac{1}{m a^2} \,, \qquad \big\langle \phi_m^{(0)}\big| \bm \nabla_{\bm u_I}\phi_n^{(1)}(\bm u)\big\rangle \sim \frac{1}{a} \,.
\end{align}
The relative corrections to the wave function is then of the order $\delta \Psi/(\phi \chi) \sim u \epsilon^3/a$, which becomes of order one for displacements $u \gtrsim a/\epsilon^3$.


\section{Application to the Migdal effect in solids}

\noindent We can apply this formalism to the determination of the Migdal effect matrix element in a solid crystal. For the sake of comparison with the results obtained via the effective theory of~\cite{Berghaus:2022pbu}, we consider a dark matter--nucleus interaction mediated by a heavy scalar mediator, corresponding to a non-relativistic Hamiltonian given by,
\begin{align}
    H_{\chi} = -\frac{g_\chi g_{\rm N}}{m_\phi^2} \sum_I \delta(\bm x_\chi - \bm X_I) \,,
\end{align}
where $\bm x_\chi$ is the dark matter position, $g_\chi$ the coupling between dark matter and the mediator, $g_{\rm N}$ that between the mediator and the nuclei, and $m_\phi$ is the mediator mass~\cite{Berghaus:2022pbu}. As already mentioned, contrary to what happens for a molecule~\cite{Blanco:2022pkt}, since the center-of-mass of the crystal is infinitely heavy, the dark matter couples to the very same coordinates appearing in the Hamiltonian in Eq.~\eqref{eq:Hsolid}.

Importantly, the Migdal effects is relevant for dark matter masses $m_\chi \gtrsim \mathcal{O}({\rm MeV})$, or, alternatively, for exchanged momenta $q \gtrsim \mathcal{O}({\rm keV})$~\cite[e.g.,][]{Berghaus:2022pbu}. The latter happens to be numerically the same as $q \gtrsim 1/a$. In this kinematical regime, the displacements probed by the dark matter are $\delta X = \epsilon \, u \lesssim 1/q$, i.e. $u \lesssim a/\epsilon \ll a/\epsilon^3$. Thus, for the range of masses of interest here, we are safely within the regime of applicability of the Born--Oppenheimer approximation, as discussed in the previous section.

\vspace{1em}

The initial state of the system comprises of the incoming dark matter, described by a plane wave with momentum $\bm p_i$, and some initial state of the solid, with quantum numbers $(n,\lambda)$. In the final state, the dark matter has momentum $\bm p_f$, and the solid will have transitioned to a state with quantum numbers $(n',\lambda')$. In order for the Migdal effect to take place, the final electronic state must be excited with respect to the initial one, i.e. $n' \neq n$. The solid wave function can be separated into adiabatic and non-adiabatic terms, as explained in the previous section. The initial and final states of the reaction then read,\footnote{From now on, we omit the explicit mention to the $\mathcal{O}\big(\epsilon^4\big)$ corrections we are neglecting, as in Eq.~\eqref{eq:Psi}.}
\begin{subequations}
\begin{align}
    |i \rangle \equiv{}& |\bm p_i, \Psi_{n\lambda} \rangle = |\bm p_i, \phi_{n} \chi_{n\lambda} \rangle + |\bm p_i, \delta\Psi_{n\lambda} \rangle \,, \\
    |f \rangle \equiv{}& |\bm p_f, \Psi_{n'\lambda'} \rangle = |\bm p_f, \phi_{n'} \chi_{n'\lambda'} \rangle + |\bm p_f, \delta\Psi_{n'\lambda'} \rangle \,.
\end{align}
\end{subequations}
The transition matrix element for a Migdal event is thus given by,
\begin{align}
    \mathcal{M} ={}& \langle f| H_{\chi} | i \rangle \equiv \mathcal{M}_{\rm ad} + \mathcal{M}_{\rm non-ad} \,,    
\end{align}
where,
\begin{subequations}
    \begin{align}
        \mathcal{M}_{\rm ad.} \equiv{}& \langle \bm p_f , \phi_{n'}\chi_{n'\lambda'}| H_\chi | \bm p_i , \phi_{n}\chi_{n\lambda} \rangle \,, \\
        \begin{split}
            \mathcal{M}_{\rm non-ad.} \equiv{}& \langle \bm p_f , \phi_{n'}\chi_{n'\lambda'}| H_\chi | \bm p_i , \delta\Psi_{n\lambda} \rangle \\
            & + \langle \bm p_f , \delta \Psi_{n'\lambda'} | H_\chi | \bm p_i , \phi_n \chi_{n\lambda} \rangle \,.
        \end{split}
    \end{align}
\end{subequations}
In a molecular context, these matrix elements correspond, respectively, to the contributions dubbed ``center-of-mass recoil'' and ``non-adiabatic coupling'' in~\cite{Blanco:2022pkt}.


\subsection{The adiabatic contribution}

The evaluation of the adiabatic part of the matrix element is rather straightforward, as it is given by,
\begin{align}
    \mathcal{M}_{\rm ad.} ={}& - \frac{g_\chi g_{\rm N}}{V\, m_\phi^2} \sum_I \int d\{\bm u\} \, e^{i \bm q \cdot\bm{X}_I} \,\chi_{n'\lambda'}^*(\bm u) \, \chi_{n\lambda}(\bm u) \notag \\
    & \times \int d\{\bm x\} \, \phi_{n'}^*(\bm x|\bm u) \, \phi_{n}(\bm x|\bm u) \,,
\end{align}
where $V$ is the volume, and $\bm q \equiv \bm p_i - \bm p_f$ is the momentum exchanged by the dark matter. The electronic wave functions are orthonormal and, since $n'\neq n$, the matrix element above vanishes, $\mathcal{M}_{\rm ad} = 0$. Therefore, in a crystal, where the center-of-mass is infinitely heavy, there is no adiabatic contribution to the Migdal effect.

For comparison, one can look at what happens in a molecule~\cite{Blanco:2022pkt}. In that instance, the dark matter couples to the actual nuclear coordinates, while the system's wave function depends on the reduced coordinates in the center-of-mass frame. This introduces a phase factor in the last line in the equation above, which prevents the integral from vanishing.


\subsection{The non-adiabatic contribution}

\noindent Before computing the matrix element associated to the non-adiabatic contribution, we start by discussing two properties of the following function,
\begin{align} \label{eq:F}
    \bm{F}_{mn}^{I} \equiv \frac{\big\langle \phi_m^{(0)}\big| \bm \nabla_{\bm u_I}\phi_n^{(1)}(\bm u)\big\rangle}{\mathcal{E}_m^{(0)}- \mathcal{E}_n^{(0)}} \,.
\end{align}
First of all, since $\mathcal{E}_n^{(0)}$ is independent on the displacements, $\bm u$, while $\phi^{(1)}_n$ is linear in them, it follows that $\bm F_{mn}^I$ does not depend on $\bm u$. Secondly, the orthogonality of the electronic wave functions means that, for $m\neq n$, 
\begin{align}
    \begin{split}
    \langle \phi_m | \phi_n \rangle ={}& \big\langle \phi_m^{(0)} \big| \phi_n^{(0)} \big\rangle \\
    & + \epsilon \left( \big\langle \phi_m^{(0)} \big| \phi_n^{(1)}(\bm u) \big\rangle + \big\langle \phi_m^{(1)}(\bm u)\big|\phi_n^{(0)}\big\rangle \right) \\
    & + \mathcal{O}\big(\epsilon^2\big) = 0 \,,
    \end{split}
\end{align}
which implies $\big\langle \phi_m^{(0)} \big| \phi_n^{(1)}(\bm u) \big\rangle + \big\langle \phi_m^{(1)}(\bm u)\big|\phi_n^{(0)}\big\rangle = 0$. By taking the gradient with respect to the nuclear displacements on both sides, one deduces that,
\begin{align}
    \bm{F}_{mn}^I = \bm{F}_{nm}^{I*} \,.
\end{align}

We also notice that the non-adiabatic correction to the wave function can also be re-written as,
\begin{align}
    \delta \Psi_{n\lambda} = \frac{\epsilon^3}{m} \sum_I \bm{\nabla}_{\bm u_I} \chi_{n\lambda}(\bm u) \cdot \sum_{m\neq n} {\bm F}_{mn}^I \, \phi_m(\bm x|\bm u) \,.
\end{align}
Indeed, by replacing $\chi_{n\lambda}^{(0)} \to \chi_{n\lambda}$ and $\phi_m^{(0)} \to \phi_m$ we make an error of order $\mathcal{O}\left(\epsilon^4\right)$, which we neglect anyways. This considerably simplifies our calculation, but one should keep in mind that the results are to be intended as only valid at lowest order in the $\epsilon$-expansion.

With this at hand, the non-adiabatic contribution to the matrix element now reads,
\begin{widetext}
    \begin{align}
        \begin{split}
            \mathcal{M}_{\rm non-ad.} ={}& - \frac{g_\chi g_{\rm N}} {V m_\phi^2} \frac{\epsilon^3}{m} \sum_{I,J} \int d\{\bm u\} \, e^{i \bm q\cdot \bm X_I} \bigg\{ \chi^*_{n'\lambda'}(\bm u) {\bm \nabla}_{\bm u_J} \chi_{n\lambda}(\bm u) \cdot \sum_{m \neq n} {\bm F}_{mn}^J \, \langle \phi_{n'}(\bm u)|\phi_m(\bm u)\rangle \\
            & \qquad \qquad \qquad \qquad \qquad \qquad \quad \; + \chi_{n\lambda}(\bm u) {\bm \nabla}_{\bm u_J} \chi_{n'\lambda'}^*(\bm u) \cdot \sum_{m\neq n'} {\bm F}_{mn'}^{J*} \, \langle \phi_m(\bm u)|\phi_n(\bm u)\rangle \bigg\} \\
            ={}& \frac{g_\chi g_{\rm N}} {V m_\phi^2} \frac{\epsilon^3}{m} \sum_I \int d\{\bm u\} \, e^{i \bm q\cdot \bm X_I} \bigg\{ i \epsilon \, \bm q \cdot {\bm F}_{n'n}^I \, \chi_{n'\lambda'}^*(\bm u) \, \chi_{n\lambda}(\bm u) - \sum_{J} \chi_{n\lambda}(\bm u) {\bm \nabla}_{\bm u_J} \chi_{n'\lambda'}(\bm u) \cdot \left[{\bm F}_{nn'}^{J*} - \bm{F}_{n'n}^J \right] \bigg\} \\
            ={}& i \frac{g_\chi g_{\rm N}}{V m_\phi^2} \frac{\epsilon^4}{m} \sum_I \int d\{\bm u\} \, e^{i \bm q \cdot \bm X_I} \, \bm q \cdot \bm{F}_{n'n}^I \, \chi_{n'\lambda'}^*(\bm u) \, \chi_{n\lambda}(\bm u) \,.
        \end{split}
    \end{align}
\end{widetext}
Specifically, to obtain the second equality, we first integrated by parts the first term, and then used the fact that $\bm{\nabla}_{\bm u_J} ( \bm q\cdot \bm X_I) = \epsilon \, \delta_{IJ} \, \bm q$.

We now notice that the function ${\bm F}^I_{n'n}$ can be rewritten in terms of the interaction potential between electrons and nuclei. First of all, we split the electronic Hamiltonian as,
\begin{align}
    H_e(\bm X, \bm x) = T_e({\bm \nabla}_{\bm x}) + V_{ee}(\bm x) + V_{e{\rm N}}(\bm x,\bm X) \,,
\end{align}
where the first terms is the electronic kinetic term, the second the electron-electron potential, and the third the electron-nucleus potential. Now, since $\phi_n$ is an eigenfunction of this Hamiltonian, we can write,
\begin{align}
    \begin{split}
        (\mathcal{E}_{n'} - \mathcal{E}_n)\langle \phi_{n'}| \bm{\nabla}_{\bm u_I} \phi_{n} \rangle ={}& \langle \phi_{n'} | [H_e, \bm{\nabla}_{\bm u_I}] | \phi_n \rangle \\
        ={}& - \langle \phi_{n'} | \bm{\nabla}_{\bm u_I} V_{e\rm N} | \phi_n \rangle \,.
    \end{split}
\end{align}
Expanding the left hand side at lowest order in $\epsilon$, and recalling Eq.~\eqref{eq:F}, one gets,
\begin{align}
    \begin{split}
        {\bm F}_{n'n}^I ={}& - \frac{1}{\epsilon} \frac{\langle \phi_{n'}|{\bm \nabla}_{\bm u_I}V_{e\rm N}|\phi_n\rangle}{\left( \mathcal{E}_{n'}^{(0)} - \mathcal{E}_{n}^{(0)}\right)^2} \\
        ={}& - \frac{\langle \phi_{n'}|{\bm \nabla}_{\bm X_I}V_{e\rm N}|\phi_n\rangle}{\omega^2} \,,
    \end{split}
\end{align}
where we used the fact that $\omega \equiv \mathcal{E}_{n'}^{(0)} - \mathcal{E}_{n}^{(0)}$ is the energy acquired by the electron due to the Migdal effect, which is also the energy released by the dark matter. We further used the relation ${\bm \nabla}_{\bm u_I} = \epsilon {\bm \nabla}_{\bm X_I}$.

Collecting everything together, the lowest order result for the non-adiabatic matrix element is,
\begin{align}
    \mathcal{M}_{\rm non-ad.} ={}& - i \frac{g_\chi g_{\rm N}}{V m_\phi^2 M \omega^2} \sum_I \int d\{\bm u\} \, e^{i \bm q \cdot \bm X_I} \chi_{n'\lambda'}^*(\bm u) \notag \\ 
    & \times \chi_{n\lambda}(\bm u) \, \bm q \cdot \langle \phi_{n'}|{\bm \nabla}_{\bm X_I}V_{e\rm N} | \phi_n \rangle \,.
\end{align}
This is precisely the matrix element found in~\cite{Berghaus:2022pbu} by means of the effective theory approach.\footnote{For a proper comparison, recall that, to preserve the unitary normalization, the nuclear wave function expressed in terms of the nuclear displacements, $\chi(\bm u)$, is related to the one expressed in terms of the nuclear coordinates, $\tilde\chi(\bm X)$, by $\chi(\bm u) = \epsilon^{3N/2} \tilde\chi(\bm X)$, where $N$ is the number of nuclei in the system.}


\section{Conclusion}

By implementing the Born--Oppenheimer approximation systematically, we computed the matrix element for the excitation of electrons in a solid, as induced by a dark matter--nucleus scattering event --- the Migdal effect. As shown also in~\cite{Blanco:2022pkt}, this approach allows to neatly separate adiabatic and non-adiabatic contributions. For a formally infinite solid, the adiabatic contribution vanishes, while the non-adiabatic one matches the result found in~\cite{Berghaus:2022pbu} via a completely independent method. Importantly, this confirms the validity of the experimental bounds determined in~\cite{SENSEI:2023zdf}, which employed the results of~\cite{Berghaus:2022pbu}.

It is interesting to note how our results highlight a clear pattern in the relative importance of adiabatic and non-adiabatic effects. For an atomic target, in fact, the Migdal effect is purely of the adiabatic type, as there is only a single nucleus~\cite[e.g.,][]{Landau:1991wop}. For a molecular target, instead, both contributions play a role, with the non-adiabatic one being more relevant for larger molecules~\cite{Blanco:2022pkt}. As a confirmation of this pattern, in a formally infinite solid, the Migdal effect is purely due to non-adiabaticities. In light of this, it would be interesting to apply the present formalism to a generic system, with an arbitrary number of nuclei, thus explicitly interpolating between the regimes discussed above. We leave this for future work.


\begin{acknowledgments}
\noindent We are grateful to Kim~V.~Berghaus, Carlos~Blanco, Benjamin~Lillard and Shashin~Pavaskar for invaluable comments and for several enlightening conversations on the nature of non-adiabatic contributions.
\end{acknowledgments}

\bibliographystyle{apsrev4-1}
\bibliography{biblio.bib}

\end{document}